\documentclass[reprint,aps,pra,floatfix,amsmath]{revtex4-2}
\usepackage{graphicx,amsmath} 

\begin{document}
\title{Transport through a lattice with local loss: from quantum dots to lattice gases}
\author{J.R.~Anglin}

\affiliation{\mbox{State Research Center OPTIMAS and Fachbereich Physik,} \mbox{Rheinland-Pf\"alzische Technische Universit\"at,} \mbox{D-67663 Kaiserslautern, Germany}}
\date{\today}

\begin{abstract}
    Recent work has studied fermion transport through a finite one-dimensional lattice of quantum dots, with localized particle loss from the central lattice site. The dots at each end of the lattice are connected to macroscopic leads, represented as zero-temperature reservoirs of free fermions with a given potential difference. Here we show how this model represents one limiting case of a larger class of models that can be realized with cold quantum gases in optical lattices. While quantum gas realizations allow many system parameters to be varied, we note limitations from finite size effects, and conclude that quantum dots and quantum gases offer complementary views on transport through lossy finite lattices.
\end{abstract}

\maketitle

\section{Introduction}
\subsection{Motivation and background}
The effects of localized loss on quantum transport are of high current interest both as examples of fundamental phenomena and as practical tools for controlling quantum states and processes \cite{H1,H2,H3,T1,T2,K1,D1,AC}. In particular, recent theoretical work \cite{AMTC0,AMTC} has investigated fermions traveling from one reservoir to another through a finite chain of sites, with one of the sites having a finite rate of particle loss. The finite number of sites means that the current is dramatically affected by reflection and discrete eigenfrequencies, and so losing particles from one site does not just drain away some current classically, but also modifies quantum interference effects. 

Reference \cite{AMTC} has solved this general problem with nonequilibrium Green’s functions in the Keldysh formalism extended to open quantum systems. The main result for the steady-state fermionic current through the finite chain is that conductance shows discrete steps, as a function of voltage between the reservoirs across the chain of sites, and that the lossy site systematically smooths out these steps, even with moderate loss rates. Higher loss rates also significantly lower the steps, reducing overall conductance. 

Letting the single lossy site be the central site, with an odd number of sites in the chain, elegantly isolates the effect of loss on the conductance steps. The conductance steps are associated with energetic access to single-particle eigenstates in the finite chain; with reflection symmetry, odd-parity eigenstates have nodes at the loss site, and transport through these channels is much less affected by the loss than transport through even-parity channels. This shows up clearly in \cite{AMTC} in conductance steps that alternate between smooth and steep.

Although \cite{AMTC} explicitly raises the possibility of experimental realizations with ultracold fermi gases in optical lattices, however, its concrete results are obtained for parameter regimes that rather represent electron transport through a chain of quantum dots, with the reservoir roles played by macroscopic wire leads. Electron tunneling between quantum dots can certainly be realized \cite{QDTun1}, and one-dimensional chains of single-fermion sites provide idealized theoretical models for transport through quantum dot arrays\cite{AMTC0,QDMod1,QDMod2}. In fact, realizing such ideally uniform chains of quantum dots is a longstanding technological challenge\cite{QDprac1,QDprac2,QDprac3}. Although microscopic control of transport in solid state systems may have important applications in the future, the simplest cases of dissipative control over transport are currently more likely to be seen in lattice-trapped quantum gases.

In this paper we therefore look critically at a quantum gas realization of this lossy transport problem. Instead of applying an open-system formalism, we replace the macroscopic fermion reservoirs of \cite{AMTC} with more lattice sites like the ones in the finite middle chain; see Fig.~1. The finite middle chain is only separated from the ``reservoir'' parts of the lattice by ``weak links'' of slower tunneling rates between sites. 

\begin{figure}\label{fig0}
\includegraphics[width=.45\textwidth]{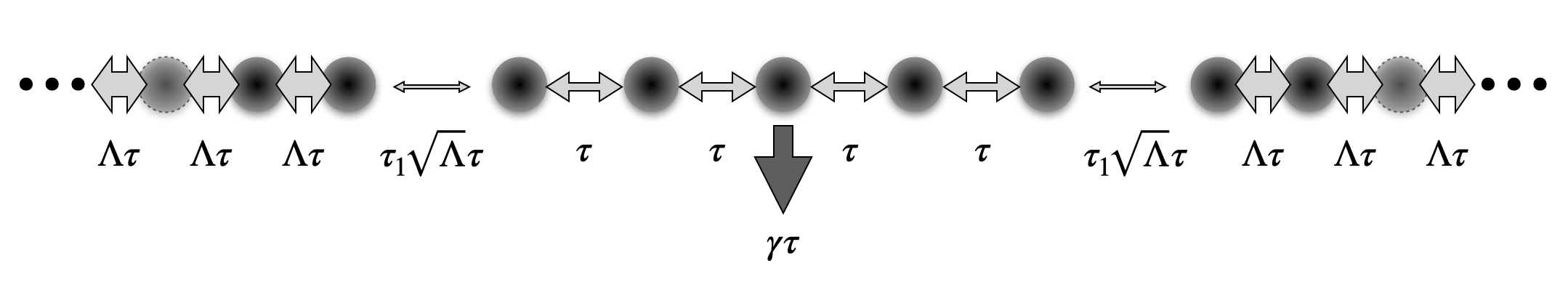}
	\caption{Spinless fermions move in a long chain of sites by nearest-neighbor tunneling; site-dependent tunneling rates define a finite chain of sites in the middle, separated by weak links from the long outer ends of the chain. Fermions can be lost from the central site. The system can model quantum dots between leads or cold atoms in an optical lattice.}
\end{figure}

We will show, first of all, that this particular class of Hamiltonian realizations of the particle reservoirs can indeed reproduce exactly the results of \cite{AMTC}. We will then re-tune our parameters to represent experiments that will likely be more straightforward with quantum gases. We will investigate both finite-temperature and finite-size effects, and conclude that experiments with quantum gases are realistically feasible, and can provide complementary insights to those offered by quantum dots.

\subsection{Paper structure}
In Section II immediately below we will introduce our lattice Fermi gas and show how all single-particle observables can be found by solving a single-particle Schr\"odinger equation. Loss from the central lattice site, which is defined through a Lindblad master equation, is shown to be described exactly by an imaginary potential term in the Schr\"odinger equation. 

In Section III we will then explain the class of initial states which let this infinite closed system exactly reproduce the observables of the open system in which a finite lattice is coupled at its ends to particle reservoirs. In particular we will present a case of our model which exactly reproduces the results of \cite{AMTC} for finite chains of quantum dots coupled to macroscopic leads.

In Section IV we will then re-tune our infinite model to correspond more closely to the most straightforward kind of quantum gas experiments, with chemical potential differences between the left and right reservoirs, but no on-site potential differences, and with densities of states in the middle and outer sections of the lattice all equal, instead of having much greater densities of states in the reservoirs. We predict steady-state currents that are qualitatively similar to those found in \cite{AMTC} for the quantum dot system with leads. There are abrupt steps in the conductance at chemical potential differences corresponding to eigenfrequencies of the isolated middle lattice; the steps are smoothed by the loss at the central site, or remain quite sharp, depending on the parity of the corresponding eigenfunctions. Although qualitatively very similar, the straightforward quantum gas scenario is found to differ in detail from the quantum dot version of the experiment.

In Section V we will consider the experimentally important constraint of finite lattice size, which in the presence of loss means that the only true steady state of the system is absence of particles (except in cases with reflection symmetry, in which odd-parity modes may be occupied, but no transport occurs). We will show, however, that if the outer ends of the lattice contain sufficiently many sites, then there is a time interval during which the system can can sustain a quasi-steady flow of particles through the middle region. We will further show that the current during this quasi-steady era corresponds closely to the steady-state current in the infinite system, with residual finite-size effects that are small for lattices with total site numbers of order 100.

We will conclude in Section VI with a summary of our results and an outlook toward further theoretical and experimental work to address bosons, finite temperatures, interactions, and multiple loss sites. A series of Appendices supply technical details and derivations of formulas provided in the main text.

\section{Reduction to a one-particle Schr\"odinger equation}
\subsection{Infinite lattices with weak links as finite lattices between reservoirs}
We consider an infinite one-dimensional lattice of sites that can be occupied by non-interacting particles that can move between sites. In second-quantized notation the Hamiltonian operator takes the form
\begin{align}\label{H}
    \hat{H}=\hbar\tau\sum_{m = -\infty}^{\infty}\left( -T_m (\hat{a}^\dagger_{m+1}\hat{a}_m + \hat{a}^\dagger_m\hat{a}_{m+1}) + V_m \hat{a}^\dagger_m\hat{a}_m\right)\;,
\end{align}
where $m$ labels each site, $\tau$ is a frequency scale, and $\hat{a}_m$ and $\hat{a}^\dagger_m$ are canonical destruction and creation operators. We will have effectively spinless fermions in mind throughout this paper, but the reduction to a single-particle problem that we are about to describe proceeds identically for bosons as well; the only difference between fermions and bosons in this problem is whether Fermi-Dirac or Bose-Einstein distributions are appropriate as initial states. Throughout this paper we will use $m$ and $m'$ to refer to sites or numbers of sites, and reserve $n$ for numbers of particles.

The site-dependent parameters $T_m>0$ and $V_m$ do not depend arbitrarily on $m$, but have a simple pattern that defines a symmetrical finite chain of $2M+1$ sites ($|m|\leq M$) centered between semi-infinite reservoir zones on the left and right, as depicted in Fig.~1: we consider hopping rates and potentials of the form
\begin{align}
    T_m &= \left\{\begin{matrix}
    1 &,& -(M+1)\leq m \leq M\\
    \Lambda &,& m < -(M+2)\;\hbox{or}\;m>M+1\\
    \tau_1\sqrt{\Lambda} &,& m = -(M+2)\;\hbox{or}\;m=M+1\end{matrix}\right.\\
V_m &=\left\{\begin{matrix}
    0 &,& |m|\leq M\\
    +\frac{V}{2} &,& m < -M\\
    -\frac{V}{2} &,& m > M\end{matrix}\right.\;.
\end{align}
Our overall frequency scale $\tau$ is thus the inter-site hopping rate in the middle chain of sites, while the dimensionless parameters $\Lambda$ and $\tau_1$ effectively define, respectively, the density of states in the infinite outer ends of the lattice, and the links between the outer ends and the finite middle section. We have included the factor of $\sqrt{\Lambda}$ with the factor $\tau_1$ because this will turn out to leave a non-trivial model in the limit $\Lambda\to\infty$. In particular this limit will let the infinite outer ends of the lattice reproduce the effects of macroscopic leads that are treated as reservoirs, when $\tau_1<1$ implies a ``weak link'' between the finite middle lattice and the outer ``reservoir'' ends.

\subsection{Localized loss at site $m=0$}
Following \cite{AMTC}, we add particle loss at site $m=0$ to our model by assuming the time evolution of the many-body density operator $\hat{\rho}(t)$ to be governed by the Lindblad master equation
\begin{equation}\label{loss}
    \frac{d}{dt}\hat{\rho} = -\frac{i}{\hbar}[\hat{H},\hat{\rho}] + \gamma\tau \left(\hat{a}_0\hat{\rho}\hat{a}_0^\dagger - \frac{\hat{a}_0^\dagger\hat{a}_0\hat{\rho}+\hat{\rho}\hat{a}_0^\dagger\hat{a}_0}{2}\right)\;.
\end{equation}
With the middle-chain hopping rate $\tau$ as our basic time scale, the dimensionless rate factor $\gamma$ defines the average rate $\gamma\tau$ at which particles are lost, in a probabilistic Poisson process, from site $m=0$. In quantum gas realizations of our model, this loss can be tuned by ejecting particles from the central lattice site using lasers or electron beams.

\subsection{Reduction to a Schr\"odinger equation with a complex potential}
\subsubsection{Single-particle observables}
Equations (\ref{H}) and (\ref{loss}) describe a many-body system of arbitrarily many non-interacting fermions or bosons. The most accessible experimental observables for many-body systems, however, are usually operators of the form $\hat{O} = \sum_{m'm}O_{mm'}\hat{a}^\dagger_{m'} \hat{a}_m$, which for appropriate matrix elements $O_{mm'}$ include particle density, momentum density, kinetic energy density, and kinetic energy current density, as well as first-order quantum coherences between arbitrary lattice sites. 

Observables of this class are called single-particle observables because they do not directly probe interparticle correlations. Their expectation values in an arbitrary many-body mixed quantum state can be given in terms of the (unnormalized)\emph{single-particle density matrix} $R_{m'm}$:
\begin{align}
    \langle \hat{O}\rangle = \mathrm{Tr}(\hat{O}\hat{\rho}) &= \sum_{m',m} O_{mm'}\mathrm{Tr}(\hat{a}^\dagger_{m'}\hat{a}_m\hat{\rho})\nonumber\\
    &=:\sum_{m',m} O_{mm'}R_{m'm}\;.
\end{align}
$R_{m'm}$ is referred to as a single-particle density matrix because, as a Hermitian matrix with non-zero eigenvalues in the Hilbert space of lattice sites $m$, it is indistinguishable from the mixed-state density matrix of a single particle, apart from a normalization factor. 

(This normalization factor is simply the average total particle number,
\begin{equation}
    \sum_m R_{mm} = \mathrm{Tr}\Big(\hat{\rho}\sum_m\hat{a}^\dagger_{m}\hat{a}_m\Big)\;.
\end{equation}
If the particles are conserved, it is often conventional to normalize $R_{m'm}$ by dividing it by its trace. Since particles can be lost in our case at site $m=0$, however, the trace of $R_{m'm}$ can in general be time-dependent. The unnormalized $R_{m'm}$ that we use is the more directly relevant quantity, anyway, for the simple $\hat{O}$-type observables in the many-body system.)

Since no information about correlations between particles is included in $R_{m'm}$, $R_{m'm}$ is in general a very incomplete description of a many-body system. It is not an approximation, however, but rather an exact representation of a restricted class of observables---which happens to include most observables of physical interest. 

\subsubsection{Single-particle master equation}
In the presence of interparticle interactions or nonlinear Lindblad operators, interparticle correlations can still affect the single-particle observables indirectly, even strongly, because time evolution couples $R_{m'm}$ to expectation values of higher-order combinations of $\hat{a}^\dagger_{m'}$ and $\hat{a}_m$ operators, in the typically infinite BBGKY hierarchy. Since our $\hat{H}$ from (\ref{H}) includes no interactions among the particles, however, and our Lindblad operator is simply $\hat{a}_0$, in our case the exact time evolution of $R_{m'm}(t)$ is given by a single-particle master equation which closes, in the sense that it involves only $R_{m'm}$ itself. Inserting our many-body Lindblad equation (\ref{loss}) into the definition of $R_{m'm}$ and applying canonical (anti-)commutation relations $\hat{a}_{m}\hat{a}^\dagger_{m'}\pm\hat{a}^\dagger_{m'}\hat{a}_{m} =\delta_{m'm}$ for either fermions or bosons reveals
\begin{align}\label{closed}
\frac{i}{\tau}\dot{R}_{m'm} =& T_{m'}R_{m'+1,m}+T_{m'-1}R_{m'-1,m}\nonumber\\
&-T_mR_{m',m+1}-T_{m-1}R_{m',m-1}\nonumber\\
&+(V_m-V_{m'})R_{m'm}\nonumber\\
&-i\frac{\gamma}{2} \big(\delta_{m'0} R_{0m}+\delta_{0m} R_{m'0}\big)\;.
\end{align}

\subsubsection{Single-particle Schr\"odinger equation}
Equation (\ref{closed}) can then be greatly simplified even further, by noting that it can be solved by a factorizing \emph{Ansatz} of the form
\begin{align}\label{Rmmt}
    R_{m'm}(t) =& \sum_{j,j'}\int d\omega\,d\omega'\, A_{j'j}(\omega,\omega')\nonumber\\
    &\times e^{i\tau(\omega'-\omega)t}\Psi^{j'*}_{m'}(\omega')\Psi_{m}^j(\omega)
\end{align}
as long as the $\Psi_{m}^j(\omega)$ are solutions $\psi_m\to\Psi_m^j(\omega)$ of time-independent single-particle Schr\"odinger equation
\begin{align}\label{1PS}
    \omega\psi_m = -T_m\psi_{m+1}-T_{m-1}\psi_{m-1} + (V_m- i\frac{\gamma}{2}\delta_{m0})\psi_m  
\end{align}
with a complex potential at $m=0$. The superscript $j$ in $\Psi^j_m(\omega)$ is used to distinguish independent solutions to (\ref{1PS}) having the same eigenfrequency $\omega$, if these exist.

In fact it can be shown that our \emph{Ansatz} (\ref{Rmmt}) includes the general solution to (\ref{closed}) as an initial value problem. See Appendix D for further discussion.

\subsection{Open or closed?}
This concludes our definition of our non-interacting lattice gas with weak links and loss. Although our system is open in the sense of having loss, the loss can be represented exactly with an imaginary potential. We do not include any external sources of particles or energy, or noise of any kind. We nevertheless propose to consider the middle chain of $2M+1$ sites, between the weak links, as an open system, for which the infinite outer ends of the lattice constitute reservoirs that can send particles into or through the middle chain of sites. In the next Section we will explain how this works. 

\section{Non-equilibrium stationary states}
\subsection{Non-equilibrium many-body system states from single-particle wave functions?}
So how do we use solutions to our single-particle Schr\"odinger equation (\ref{1PS}) to describe non-equilibrium steady states as in \cite{AMTC}, with particle transport between the two reservoirs through the finite middle chain with its single lossy site?

We will consider special cases of (\ref{Rmmt}) of the form
\begin{equation}\label{Rmmstat}
    R_{m'm} = \sum_{j=L,R}\int\!d\omega\,D_j(\omega)f_j(\omega)\Psi^{j*}_{m'}(\omega)\Psi_m^j(\omega)\;,
\end{equation}
with $j=L,R$ denoting wave functions that are incident from the left and right, respectively. Here 
$D_j(\omega)$ are the corresponding densities of states, and $f_j(\omega)$ are Fermi-Dirac distributions with independent temperatures and chemical potentials for $j=L$ and $j=R$. (In future work it will be straightforward to use Bose-Einstein distributions instead, and describe analogous transport problems with quantum Bose gases. In our concluding Section VI we suggest that the bosonic problem will be less interesting, however, unless interactions are present.) 

Since in general we allow a potential difference $V$ between the left and right ends of our infinite lattice, as well as different equilibrium distributions for $f_L$ and $f_R$, our stationary states will include transport in general.

In Appendix A we will explicitly construct the $\Psi^{L,R}_m(\omega)$ solutions to (\ref{1PS}); in the remainder of our main text we will use reflection and transmission coefficients that are derived in that Appendix. Here we explain why an $R_{m'm}$ of the form (\ref{Rmmstat}) really does represent non-equilibrium steady states of our system, and then outline how we will use these $R_{m'm}$ to compute observable quantities in these states.

\subsection{Left- and right-incident normalized eigenfunctions}
The solutions $\Psi_m^{L,R}(\omega)$ that we construct explicitly in Appendix A have the following ``scattering'' forms for $|m|>M$:
\begin{align}\label{LRinc}
    \sqrt{2\pi}\Psi_m^L(\omega) &=\left\{\begin{array}{ccc}
       e^{ik_Lm}+\mathcal{R}_L(\omega)e^{-ik_Lm} &\;,\;  &m < -M  \\
       \sqrt{\frac{sin(k_L)}{sin(k_R)}}\mathcal{T}(\omega) e^{ik_R m}  & \;,\;  & m > M
    \end{array}\right.\nonumber\\
    \sqrt{2\pi}\Psi_m^R(\omega) &=\left\{\begin{array}{ccc}
       \sqrt{\frac{sin(k_R)}{sin(k_L)}}\mathcal{T}(\omega) e^{-ik_L m} &\;,\;  & m < -M  \\
        e^{-ik_Rm}+\mathcal{R}_R(\omega)e^{ik_Rm} & \;,\;  & m > M
    \end{array}\right.\;,
\end{align}
for transmission and reflection coefficients $\mathcal{T}$ and $\mathcal{R}_{L,R}$ that we compute in Appendix A and will use below. The behaviors of $\Psi^{L,R}_m$ for $|m|\leq M$ are shown in Appendix A, but will not be needed for any results we will present in our main text.

Since we have neither time-reversal nor (unless $V=0$) spatial reflection symmetry, the fact that the transmission coefficient $\mathcal{T}(\omega)$ is the same for both left- and right-incident waves is not obvious, but it turns out to be so in all possible cases of our model. The reflection coefficients $\mathcal{R}_{L,R}(\omega)$, in contrast, do not coincide in general. In the special cases that we will consider explicitly within our main text, however, $\mathcal{R}_{L,R}\to\mathcal{R}(\omega)$ \emph{are} in fact the same, because $k_{L,R}$ are the same, because these cases will either have $V=0$ or $\Lambda\to\infty$.
 
In the $|m|>M$ regions the wave numbers $k_{L,R}(\omega)$ can be identified immediately, by recognizing that plane waves $e^{\pm i k_{L} m}$ are local solutions to the single-particle Schr\"odinger equation (\ref{1PS}) for $m<-M$, as are $e^{\pm i k_{R} m}$ for $m>M$, if
\begin{equation}\label{DOS1}
    -2\Lambda\cos(k_{L,R}) \pm \frac{V}{2} = \omega 
\end{equation}
where the $+(-)$ branch of $\pm$ applies for $L(R)$. The ratios of $\sin(k_{L,R})$ that appear as factors before $\mathcal{T}_{L,R}$ are the usual ratios of group velocities $d\omega/dk$ that appear in one-dimensional transmission and reflection problems when the left and right asymptotic group velocities may be different, as here if $V\not=0$. We can also note from (\ref{DOS1}) that the densities of states which appear in our stationary $R_{m'm}$ integral (\ref{Rmmstat}) are
\begin{equation}\label{DOS3}
    D_{L,R}(\omega) = \frac{1}{\frac{d\omega}{dk_{L,R}}}=\frac{1}{2\Lambda\sin\big(k_{L,R}(\omega)\big)}\;,
\end{equation}
which as usual in one-dimensional systems are the inverse group velocities.
\subsection{Why scattering wave functions?}
The reason for defining these left- and right-incident wave functions $\Psi_m^{L,R}$ in this way (\ref{LRinc}) is that they allow us to describe left and right reservoirs with different temperatures or chemical potentials, by considering separate grand canonical populations of the second-quantized $L$ and $R$ modes.

It may just seem obvious that uncorrelated populations of second-quantized $\Psi^L_m$ and $\Psi^R_m$ modes represent fermions approaching our middle chain from different equilibrium conditions. In fact the subject is worth some discussion. The set of all $\Psi_m^L$ and $\Psi_m^R$ for all $\omega$ together (plus any discrete bound states---see Appendix D) is complete, so any state can be represented in terms of them; but superpositions of $\Psi_m^L$ and $\Psi_m^R$ can also provide a complete set of states, and in general these superpositions will imply correlations and even coherences between incident particles from the left and incident particles from the right. It may be a physically natural condition to rule out such correlations, but it is still a non-trivial condition.

We can confirm in the following way that the kind of non-equilibrium steady state that we wish to consider in $R_{m'm}$ is indeed the sum of a grand canonical distribution over $\psi_m^L$ modes and a second grand canonical distribution over $\psi_m^R$ modes. Suppose that we prepare the initial state of our infinite system by starting with $\tau_1\to0$, so that all $\mathcal{T}(\omega)\to 0$ and all $|\mathcal{R}_j(\omega)|\to 1$. The left and right ends of the lattice are now completely separate semi-infinite uniform lattices, and the set of all $\Psi_m^L$ modes is a complete set of second-quantized normal modes in the left semi-infinite lattice, while the set of $\Psi_m^R$ modes is complete in the right. 

We throw infinitely many particles into both of these separate semi-infinite lattices, by populating the $\Psi_m^L$ and $\Psi_m^R$ modes in second quantization. We then wait as long as it takes for the particles to reach equilibrium, either through contact with external baths which will later remove, or through interactions among the particles which are weak enough to have negligible effects on the nature of the steady state but which will thermalize the system over long enough times.

We now slowly raise $\tau_1$ to its desired steady-state value, so slowly that the $\Psi_m^{L,R}$ will all adiabatically evolve into their finite-$\tau_1$ forms, while keeping their average occupation numbers invariant. Over this long adiabatic ramping of $\tau_1$, whatever finite initial population was in the finite middle chain will either be lost at the impurity or else escape away to infinity through the weak links, while the infinite populations in the $\Psi_m^{L,R}$ so that the population remaining in finite $m$ sites is entirely given by the two grand canonical populations of $\Psi_m^{L,R}$, as assumed in (\ref{Rmmstat}).

Realistic state preparations may not take infinite time, or involve literally infinite systems, but they are quite likely to conform qualitatively to this adiabatic scenario. In any case, the steady state that results from the adiabatic $\tau_1$-ramping scenario must clearly be the kind of state that we mean when we speak of steady flow between reservoirs at two different energies and chemical potentials. If there is any other kind of state preparation that will produce such a state, the state itself will still be the same, up to the small errors that are always involved in treating large but finite systems in quasi-steady states as if they were literally in equilibrium.

\subsection{Cases with imaginary $k_{L,R}$}
In general it may happen that one or the other of $k_{L,R}(\omega)$ as given by (\ref{DOS1}) is complex. No cases with complex $k_{L,R}$ will need to be considered here in our main text, but for possible use in future work we discuss this technicality further in Appendix B.

\subsection{Observables in the steady state}
From our steady-state single-particle density matrix (\ref{Rmmstat}), and the explicit eigenfunctions we derive in Appendix A, it is straightforward to compute several steady-state observables as functions of the Hamiltonian parameters and the initial states of the left and right lattice ends. The most basic observable is the local occupation number at each site $m$, which is simply a diagonal element of the unnormalized density matrix:
\begin{equation}
    \bar{n}_m = \mathrm{Tr}(\hat{\rho}\hat{a}^\dagger_m\hat{a}_m) = R_{mm}\;.
\end{equation}
Within the finite middle chain of sites, $\bar{n}_m$ for $|m|\leq M$ will depend on the $\Psi_m^{L,R}(\omega)$ for $|m|\leq M$, and will therefore not have any simple relation to the $\mathcal{T}$ and $\mathcal{R}_{L,R}$ coefficients that define the wave functions for $|m|>M$. We will focus instead on observables that relate directly to transport.

The simplest of these is the particle current. From the single-particle master equation (\ref{closed}) we obtain
\begin{align}\label{cont}
    \frac{d}{dt}\bar{n}_m &= -\gamma\tau\delta_{m0}\bar{n}_0-(J_{m+1}-J_{m})\nonumber\\
    J_m &= i \tau T_{m-1}(R_{m,m-1}-R_{m-1,m})\;.
\end{align}
Since the first line of (\ref{cont}) is an obvious discrete version of the continuous continuity equation, we recognize $J_m$ as the particle number current from left to right, which is also proportional to the average momentum at site $m$. 

In a stationary state, $\bar{n}_m$ is time-independent, and so we conclude that the particle number current is piecewise constant as a function of $m$:
\begin{equation}
J_m = \tau
\left\{\begin{matrix}
    I+\frac{\gamma\bar{n}_0}{2} &\;,\;& m\leq 0\\
    I-\frac{\gamma\bar{n}_0}{2} &\;,\;& m > 0\;,
\end{matrix}\right.
\end{equation}
for constant average current $I$. Since the difference between the piecewise constant currents on each side of $m=0$ thus represents particle transport \emph{into} the central lossy site from the two reservoirs, we can follow \cite{AMTC} and take $I$ as a representation of net transport \emph{through} the finite middle lattice from the left reservoir into the right reservoir. With our stationary $R_{m'm}$ represented in the form (\ref{Rmmstat}), we can evaluate $J_{L,R}$ by evaluating $J_{m}$ for $|m|>M$, and then use our $|m|>M$ forms for $\Psi_m^{L,R}$ in (\ref{LRinc}) to conclude
\begin{equation}\label{Iint}
    I = \frac{1}{4\pi}\int\!d\omega\,\Big(f_L(\omega)-f_R(\omega)\Big)\Big(1+|\mathcal{T}|^2-|\mathcal{R}|^2\Big)\;,
\end{equation}
where $f_{L,R}(\omega)$ are equilibrium occupation distributions that can be different for the two ends of the lattice. In (\ref{Iint}) we have assumed $\mathcal{R}_{L,R}\to\mathcal{R}$ as they do in the cases $V=0$ or $\Lambda\to\infty$; a more general formula for $I$ is given in Appendix C. The density of states factors $D_{L,R}(\omega)$ from (\ref{Rmmstat}) have not been omitted accidentally from (\ref{Iint}): they are truly absent, being conveniently canceled by group velocity factors that appear in $J_m$.

Additional single-particle observables that may be of interest, and that can probably be measured in quantum gas experiments, include the local energy at site $m$, which is also simply related to the momentum current, as well as the energy current, which could offer insight into how the central lossy cite absorbs heat as well as particles. We will only consider the particle current $I$ here in our main text, but we derive the corresponding formulas for the other observables in Appendix C.

\subsection{Semi-infinite lattices as macroscopic leads: The $\Lambda\to\infty$ limit}
We now confirm that our infinite lattice model, with weak links between the finite middle chain and the semi-infinite ends, does indeed represent a finite middle chain coupled to particle reservoirs.

\subsubsection{Transmission and reflection coefficients $\mathcal{T}$ and $\mathcal{R}$ for $\Lambda\to\infty$}
As we show in Appendix A, the dimensionless parameter $\Lambda$ does not appear explicitly anywhere in the expressions for $\mathcal{R}_{L,R}$ and $\mathcal{T}$. $\Lambda$ decisively affects $k_{L,R}(\omega)$ according to (\ref{DOS1}), however, and $k_{L,R}$ do both appear in $\mathcal{R}_{L,R}$ and $\mathcal{T}$. In the limit $\Lambda\to\infty$, in particular, the complicated general formulas of Appendix A for $\mathcal{R}_{L,R}$ and $\mathcal{T}$ simplify dramatically.

Since the outer ends of the lattice are just large, uniform lattices, propagating modes with real $k_{L,R}$ can have frequencies within the continuous bands $|\omega \mp V/2|\leq 2\Lambda$, respectively for the left and right ends. For large $\Lambda$, these bands become very wide compared to the range of frequencies $|\omega|\leq 2$ within which the finite middle chain of sites can support real $k$ modes.  Modes with $|\omega|\gg 2$ can propagate in the outer ends, but they must decay so sharply within $|m|< M$ that they contribute negligibly to any observables within the middle chain, including the particle current through the middle chain. As far as any such observables are concerned, therefore, we can certainly restrict our attention to $|\omega|\ll \Lambda$ when $\Lambda$ is large. For all these $|\omega|\ll\Lambda$, then, the limit $\Lambda\to\infty$ means that $k_{L,R}(\omega)$ are both negligibly different from $\pi/2$.

This lets us simplify our generally complicated formulas by replacing $k_{L,R}\to\pi/2$. The physical meaning of this mathematical convenience is that that for infinite $\Lambda$, the density of states $1/(dk_{L,R}/d\omega)$ in the ``reservoir'' ends of the lattice is uniform over the entire frequency range that matters for transport through the finite middle chain. This reproduces the uniform density of states that was assumed for the external leads in \cite{AMTC}, and this can be considered the ultimate reason why the $\Lambda\to\infty$ limit lets our infinite lattice model duplicate the open system.

With the replacement $k_{L,R}\to\pi/2$, then, our expressions from Appendix A simplify---in particular $\mathcal{R}_{L,R}\to\mathcal{R}$ because $k_L = k_R$. Defining $k(\omega)$ without subscript to be the wave number in the middle chain,
\begin{equation}\label{kom}
     k(\omega) = \cos^{-1}\Big(-\frac{\omega}{2}\Big)\;,
\end{equation}
we obtain (see Appendix A)\begin{widetext}
\begin{align}\label{transmit1}
1+|\mathcal{T}|^2-|\mathcal{R}|^2 =&\frac{2}{|Z(\omega)|^2}\left[\sin^2(k)+\frac{\gamma}{4}\left(\frac{\sin^2[k(M+1)]}{\tau_1^2}+\tau_1^2\sin^2(kM)\right)\right]\nonumber\\
|Z(\omega)|^2 =& \left(\frac{\sin[k(M+1)]\cos[k(M+1)]}{\tau_1^2}+\tau_1^2\sin(kM)\cos(kM)\right)^2\nonumber\\
&+ \left[\sin(k)+\frac{\gamma}{4\sin(k)}\left(\frac{\sin^2[k(M+1)]}{\tau_1^2} + \tau_1^2\sin^2(kM)\right)\right]^2\;.
\end{align}\end{widetext}

Since we consider small $\tau_1$, to make weak links between the outer ends and the middle chain, the important feature in (\ref{transmit1}) is that the denominator $|Z(\omega)|^2$ is large (of order $\tau_1^{-4})$, making the current contribution from a given $\omega$ small, unless either $\sin[(M+1)k(\omega)]$ or $\cos[(M+1)k(\omega)]$ is small. These are precisely the conditions $\sin[2k(\omega)(M+1)]=0$ that define the $2M+1$ discrete eigenfrequencies
\begin{align}
\omega_n &= -2\cos(k_n)\nonumber\\
k_n &=\frac{n\pi}{2(M+1)}\quad ,\quad n\in [1,2M+1]
\end{align}
of the middle chain when it is isolated ($\tau_1\to0$ so that the outer ends of the lattice can be ignored). Transport through the weakly linked middle chain will therefore be small except for particles with frequencies close to one of these resonant frequencies $\omega_n$.

It can further be noted, however, that the numerator in $1+|\mathcal{T}|^2-|\mathcal{R}|^2$ in (\ref{transmit1}) can also somewhat large---of order $\gamma\tau_1^{-2}$---unless $\sin[(M+1)k(\omega)]$ is small. For moderate $\gamma$ this broadens the transmission resonances at $\cos[(M+1)k_n]=0$, although for large $\gamma$ the denominator term of order $\gamma^2\tau_1^{-4}$ also lowers the contributions of these resonances. 

At the resonances with $\sin[(M+1)k_n]=0$, in contrast, all the terms with negative powers of $\tau_1$ vanish, and small $\tau_1$ makes the transmission factor $(1+|\mathcal{T}|^2-|\mathcal{R}|^2)/2$ close to one. This is easy to understand: the isolated-chain eigenfrequencies $\omega_n$ with $\sin[(M+1)k_n]=0$ are those for which the isolated-chain eigenfunctions are odd functions of $m$ and so have nodes at the lossy site $m=0$. Transport through these channels is protected against loss by parity symmetry, as reported in \cite{AMTC}. Transport at their resonant frequencies is only affected by loss insofar as the weak links provide non-resonant coupling to other modes of the finite chain that do not have nodes at $m=0$ and so do have loss. 

\subsubsection{Current $I$ through the chain for $\Lambda\to\infty$}
To confirm that we are reproducing the results of \cite{AMTC}, we follow it in considering fermions at zero temperature, with a non-zero voltage difference $V$ ranging through the finite chain bandwidth $-2<\omega<2$. Since the density of conduction-band electrons in a metal lead is a nearly constant property of the material regardless of electrostatic potential, we keep the chemical potentials (\textit{i.e.} Fermi energies) $\mu_{L,R}$ in the left and right reservoir $f_{L,R}(\omega)$ equal to the voltages $\pm V/2$, so that the reservoirs differ in potential but have equal fermion densities:
\begin{equation}\label{fLR}
    f_{L,R}(\omega) = \theta\big(\pm\frac{V}{2}-\omega\big)\;.
\end{equation}
\begin{figure}\label{fig:AMTCcomp1}
\includegraphics[width=.45\textwidth]{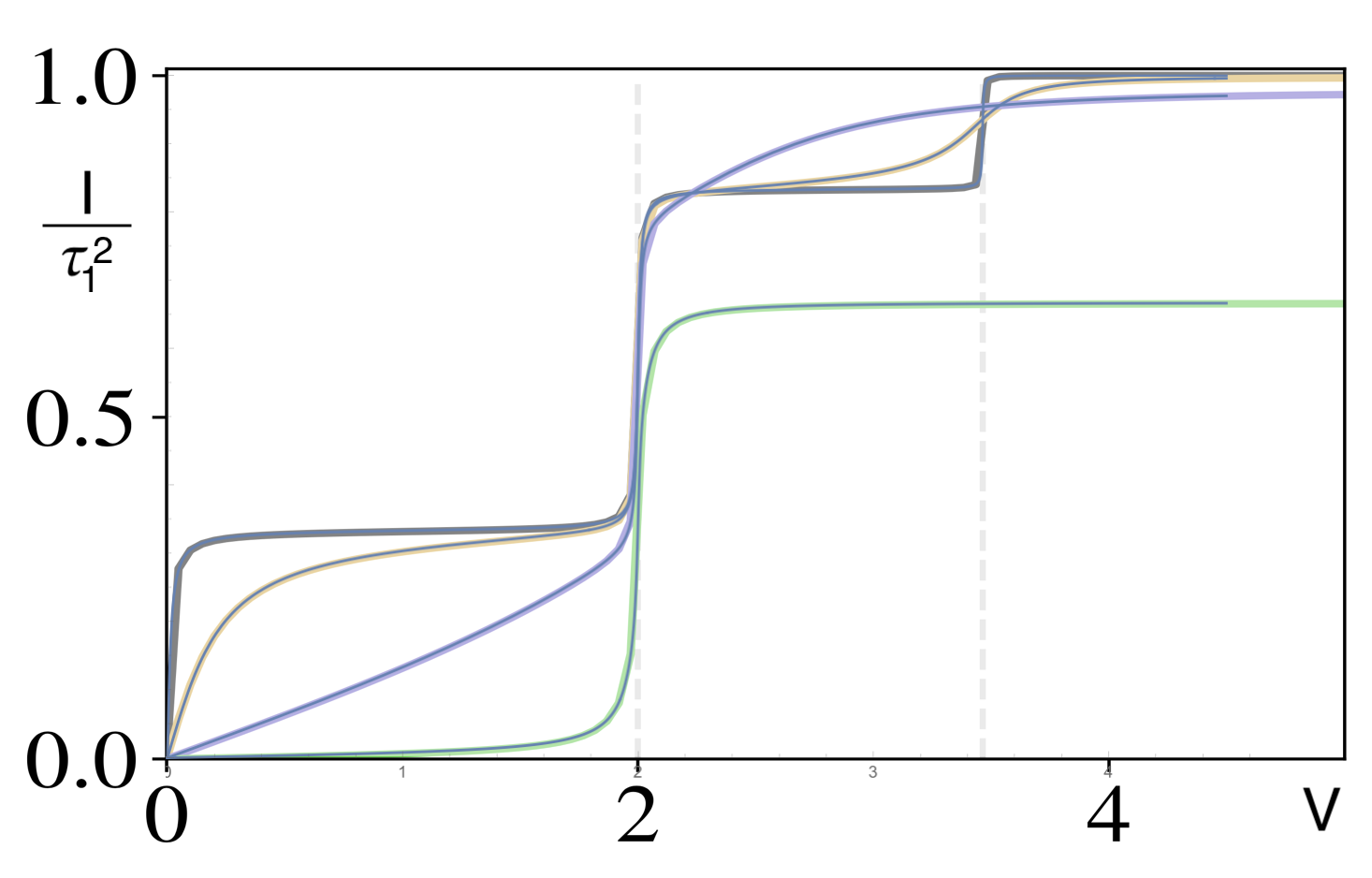}
	\caption{The dimensionless current $I$ through a finite middle chain of five sites ($M=2$), divided by $\tau_1^2$ as in \cite{AMTC}, as a function of potential difference $V$ between the left and right semi-infinite ends of the infinite lattice, for four different values of $\gamma$ (from top to bottom at the left side of the plot: $\gamma = 0$, $0.5$, $5$, $100$.) In all cases $\tau_1=0.1$. The thin, dark curves that only extend to $V=4.5$ are computed by numerical integration of our (\ref{Iint}) with (\ref{transmit1}) and (\ref{fLR}), while the broader, fainter curves that extend to the right edge of the plot are taken from Figure 5a) of Reference \cite{AMTC}. The thin, dark curves lie directly on top of the broad, faint curves, confirming the equivalence of the present model for $\Lambda\to\infty$ and the open system treated in \cite{AMTC}.}
\end{figure}
As we see in Fig.~2, our infinite lattice model exactly reproduces the results from \cite{AMTC} for a finite lattice between two macroscopic leads, treated with Kheldysh path integrals.  Figure~2 differs only in plotting resolution from Figure 5 of \cite{AMTC}. Scaling the lattice weak link as $\tau_1\tau\sqrt{\Lambda}$ while the outer lattice ends have hopping rate $\Lambda\tau$, and then taking $\Lambda\to\infty$ to make the reservoir density of states effectively constant, maps the infinite lattice gas model onto the macroscopic leads problem of \cite{AMTC}.

\section{A quantum gas version}
\subsection{The quantum gas set-up}
Having confirmed that the infinite lattice model with weak links can indeed describe the finite chain of quantum dots coupled to leads as studied in \cite{AMTC}, we now examine a version of the lattice problem that would be more straightforward to realize with an ultracold Fermi gas in a one-dimensional optical lattice (or a three-dimensional array of many one-dimensional lattices, allowing single-shot measurements of ensembles of the one-dimensional system). While the limit of large $\Lambda$ could be achieved by making $\tau$ small, with high potential barriers between the sites in the middle chain, this would slow the system's approach to a steady state, possibly to the point where steady-state behavior could not be seen within the sample lifetime. We therefore focus on the experimentally more straightforward case $\Lambda=1$.

The on-site potential in the lattice can also be made non-uniform in cold atom experiments, but it may be challenging to make the potential piecewise uniform with abrupt steps between the middle and the outer ends of the lattice. We therefore set $V\to0$, leaving the two weak links as the only nonuniformity in the lattice. With no analogous constraint to the fixed fermion density of metal leads, however, we are free to let the chemical potentials $\mu_{L,R}$ in the two lattice ends be arbitrary. Varying $\mu_L-\mu_R$ while keeping $V=0$ will turn out to have similar effects to varying $V$ with Fermi energies locked to $\mu_{L,R}=\pm V/2$; the mechanism of opening more transmission channels as the Fermi level rises is exactly the same, after all. 

With no need to maintain fixed Fermion densities in both reservoirs, moreover, we are free in the quantum gas scenario to let $\mu_R$ lie below the ground state energy, so that no particles at all are incident from the right, and the entire current is from left reservoir to right reservoir, through the middle chain with its central site loss, minus reflection from the weak links back into the left reservoir. This means that instead of (\ref{fLR}) we now take
\begin{equation}\label{fLRnew}
    f_L(\omega) = \theta(\mu_L-\omega)
\end{equation}
and $f_R=0$.

This less symmetrical but simpler set-up also has the advantage of showing more conductance steps. In the $\pm V\not=0$ scenario with equal fermion densities in leads, the currents from right and from left exactly cancel in the frequency range $\omega < -V/2$, so that the integral (\ref{Iint}) over $\omega$ is effectively over the symmetrical middle range $|\omega|<V/2$. Since the resonant frequencies of the transmission channels through the finite middle chain are in symmetrical pairs $\pm\omega_n$, raising $V$ effectively only opens new channels in pairs, producing only $M+1$ conductance steps. In the quantum gas scenario with particles coming only from the left reservoir, in contrast, raising $\mu_L$ through the system's frequency band sweeps through each of the $2M+1$ transmission resonances separately.    
\begin{figure}\label{QGversion}
\includegraphics[width=.45\textwidth]{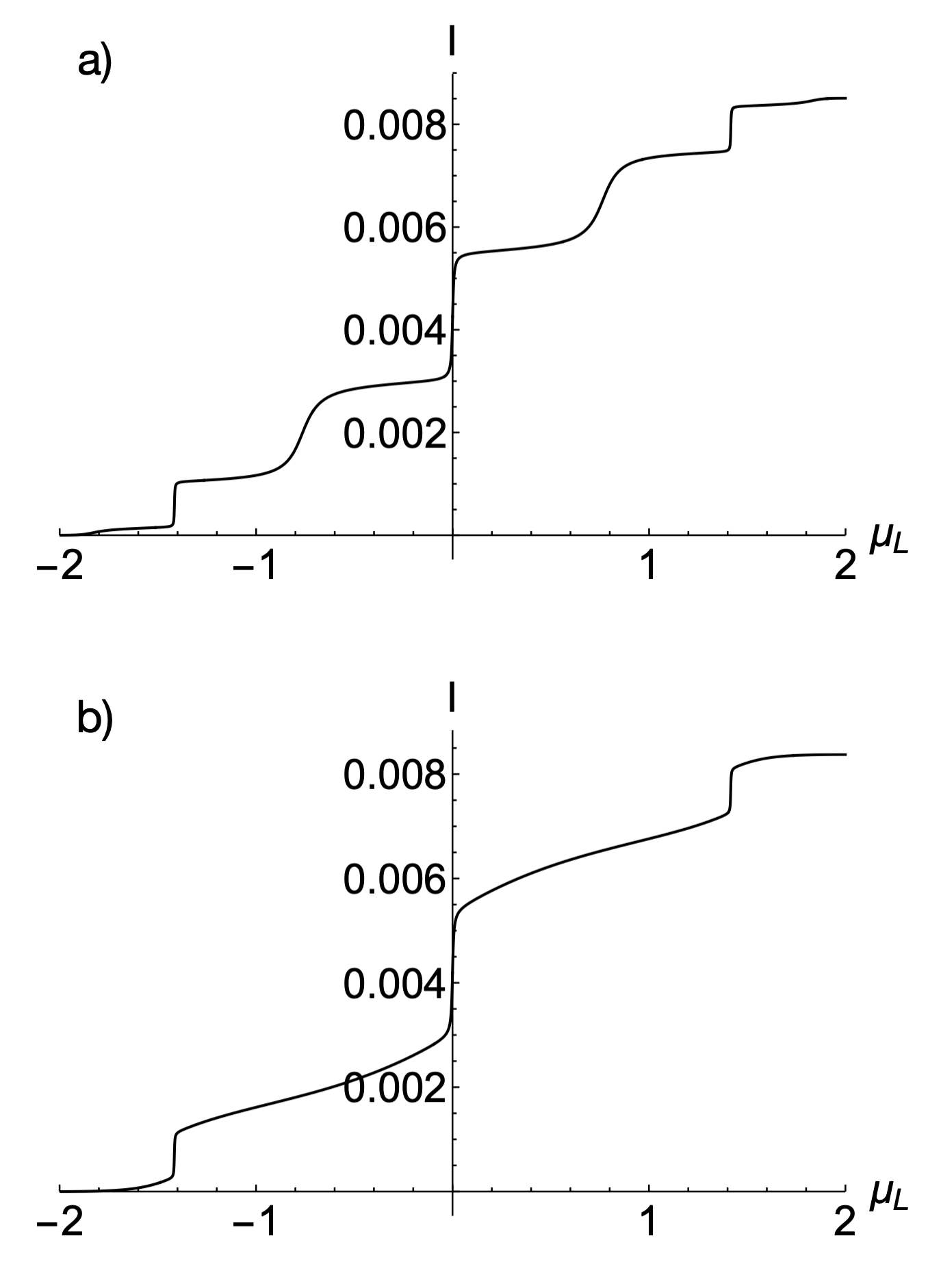}
	\caption{Dimensionless fermion current $I$ at zero temperature, as a function of left-side chemical potential $\mu_L$, for two different central-site loss rates $\gamma$. Both plots show $\tau_1=0.1$ as in Fig.~2, but with $V=0$, $\Lambda=1$, and no fermions incident from the right. Although the middle chains here have seven sites ($M=3$), the leftmost and rightmost current steps are small and gentle in a), and completely flattened in b), leaving only five current steps clearly visible. In a) we have $\gamma=0.5$, while b) shows $\gamma=5$. As well as suppressing the outermost of the seven current steps, the moderate loss in a) has noticeably smoothed out two of the five more visible steps in the current, while the stronger loss in b) has smoothed them completely into gradual slopes. The other three visible current steps are not significantly affected by loss in either case, since they are due to transport through odd-parity channels having nodes at the lossy site $m=0$. Although $I(\mu_L)$ differs in detail from analogous plots of $I(V)$ for quantum dots between leads, the same phenomena of conductance steps with parity-dependent smoothing by loss are revealed.}
 \end{figure}

Figure~3 shows two quantum gas analogs of Fig.~2, plotting the dimensionless steady-state current $I$ as a function of left-side chemical potential $\mu_L$, with no fermions incoming from the right. The current plotted in Fig.~3 is the same integral (\ref{Iint}) as in Fig.~2, but with the different $f_{L,R}(\omega)$ of (\ref{fLRnew}) as well as the more complicated $\mathcal{R}$ and $\mathcal{T}$ coefficients that are given by inserting $V\to0$ and $\Lambda\to1$ in the general formulas of Appendix A.

Five steplike rises in conductance through the finite middle chain of seven sites ($M=3$) are clearly visible in Fig.~3; looking closely at a), one can also discern two more small, smooth steps at the left and right edges of the plot. Also apparent is the effect of loss in making some steps noticeably steeper than others. In Fig.~3a) the tiny first and seventh steps, as well as the more visible third and fifth steps, are all due to even-parity channels that are affected by the loss at $m=0$, and so more gently sloped than the three other steeper steps, which are due to odd-parity channels with nodes at $m=0$. The ten-times-higher loss rate in Fig.~3b) is enough to suppress the first and seventh steps entirely, and eliminate the shoulders of what were the third and fifth steps in a), leaving them as gradual slopes, while still leaving the three parity-protected transport resonances as sharp steps.

Many different parameter regimes of this kind of lossy transport can be investigated in quantum gas experiments, which are highly tunable. While quantum gases in optical lattices may in this sense be a more flexible platform than quantum dots, quantum gases do have their own constraints. While room temperature can easily be effectively zero temperature for conduction electrons in metals, even nanokelvin temperatures may not be negligible for ultracold Fermi gases. And while macroscopic leads are effectively infinite in size, optical lattices in experiments are bound to be finite. We will examine finite temperature effects next, and treat the conceptually greater difference of finite system size in the following Section.

\subsection{Finite temperature effects}
Finite temperature effects are incorporated straightforwardly by replacing the step function form (\ref{fLR}) of the zero-temperature Fermi-Dirac function with the general case
\begin{equation}
    f_L(\omega)=\frac{1}{e^{\frac{\omega+\mu_L}{T}}+1}
\end{equation}
for temperature $T_L\to T$, leaving $f_R(\omega)=0$. The effect of finite temperature is likewise simple: it smears out sharp features of frequency. 

Since the main phenomena to be studied in this kind of transport are the conductance steps and their parity- and $\gamma$-dependent steepness, thermal smearing of the conductance curve is a serious issue for experiments. Since the frequency spacing of our transmission resonances is of order $2/(2M+1)$, temperatures higher than this spacing (in units of $\hbar\tau/k_B$) will wash out the conductance steps. Even if $T/\tau$ is below the $2/(2M+1)$ threshold, moreover, thermal smoothing of the conductance steps can reduce their dramatic signature of discrete transmission channels to only a slight waviness in $I(\mu_L)$, as seen in Fig.~4a); in this regime, moreover, the additional smoothing of even-parity channels due to the $m=0$ loss can be hard to discern. Both the steps and the effect of parity can still be seen at sufficiently low temperatures, as in Fig.~4b), but such low temperatures may be experimentally challenging. If the basic $T< 2\tau/(2M+1)$ limit can be beaten, the problem of losing visibility of the parity effect for $T\lesssim \gamma$ can be solved by increasing $\gamma$, as in Fig.~4c).
\begin{figure}\label{fig:FT1}
\includegraphics[width=.45\textwidth]{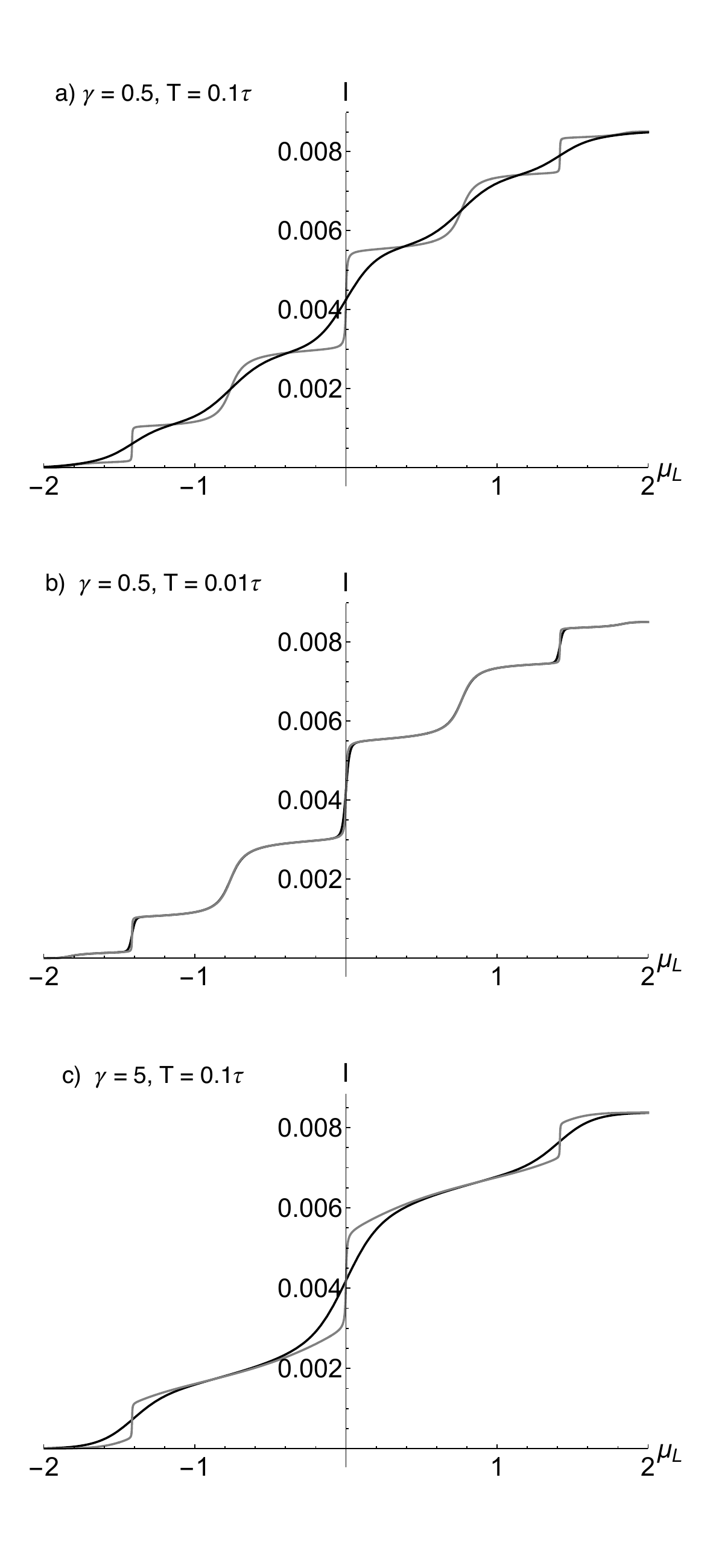}
	\caption{Ultracold Fermi gas current $I$ through the lossy impurity, as a function of left-side chemical potential $\mu_L$, at finite left-side temperature (thin black curves), with zero-temperature results for comparison (thick gray curves). All three plots show $V=0$, $\tau_1=0.1$, and $M=3$ as in Fig.~3 (seven sites in the middle chain, although at most five current steps are large enough to see clearly), with no fermions incoming from the right. In a) we have $\gamma=0.5$ and $k_BT_L=0.1\hbar\tau$ (black curve); the comparatively high temperature smooths out all the abrupt zero-temperature steps into slopes that are similarly gradual. In b) we lower the temperature to $k_BT=0.01\hbar\tau$; the steepest zero-temperature steps are only slightly less steep at this low temperature, and the effect of loss in smoothing out the second and fourth visible steps is still noticeable. In c) we raise the loss rate $\gamma$ to 5; with stronger loss, the same temperature $k_BT=0.1\hbar\tau$ as in a) no longer buries the smoothing effect of the loss on the second and fourth visible steps.}
 \end{figure}

In condensed matter systems with Fermi temperatures in thousands of Kelvin, even room temperature is negligibly different from zero temperature. In quantum gas experiments, in contrast, the ratio $k_BT/(\hbar\tau)$ of temperature to hopping rate cannot be made arbitrarily low. While it is easy to make $\tau$ small by increasing laser power in the optical lattice to reduce tunnelling rates exponentially, reducing the lattice strength in order to accelerate tunnelling will also weaken the tight binding approximation and turn on excitations out of the lowest band, so that our single-band lattice model (1) breaks down. A realistic upper bound on $\tau$ is probably in the range of $10^4$ s$^-1$, corresponding to a temperature range in hundreds of nanoKelvin\cite{HM}.  
With the currently achievable temperatures for lattice fermions in tens of nK, reaching $k_BT/(\hbar\tau)<0.1$ will likely require future technical progress. As Fig.~4 shows, a lower ratio like $k_BT/(\hbar\tau)=0.01$ may be needed to compete with the current step sharpness of quantum dot experiments, but a ratio of $0.1$ can still be sufficient to show the steps as well as the smoothing effect of loss in dependence on parity. As in many potentially interesting cold atom experiments, reaching significantly lower temperatures will be an important challenge.

We now turn to the issue of finite size effects in quantum gas realizations, which requires some new concepts but probably poses less serious problems than finite temperature.

\section{Finite size effects}

\subsection{Finite size and finite time}
Cold atom experiments can flexibly tune many parameters, but the number of lattice sites cannot really be tuned to infinity. Although the finite sizes of cold atom ``reservoirs'' can limit the effects that can be seen in cold atom experiments, we will now show how measurements of finite time evolution can in fact reproduce the steady-state behavior of infinite systems.

In the finite system with a lossy impurity there can be no steady state current: currents can only be transient, strictly speaking, because in the limit of infinite time all initial particles will have simply been lost. Technically transient currents do not have to be short-lived phenomena, however. They can easily be quasi-steady over intermediate time scales. 

\subsection{Quasi-stationary currents}
For example, a natural initial state to prepare experimentally is a finite-system analog of the current-from-the-left scenarios shown for infinite lattices in Fig.~3. We can prepare the initial state in the same way that we described above for the infinite system, by initially setting $\tau_1\to0$ to fully break the weak link and isolate the middle chain of our lattice from its outer ends. We then populate the isolated left end of the lattice with fermions in equilibrium, leaving the middle and right-end lattices empty. Since adiabatically ramping up $\tau_1$ would still just lose all of our finite number of particles, however, we instead abruptly restore non-zero $\tau_1$ $t=0$, and let the left-end population of fermions flow through the weakly linked middle chain, with its loss at $m=0$, and into the right lattice end. At different finite times $t>0$ we measure the decreasing population difference between left and right sides,
\begin{equation}
    \frac{\Delta \mathcal{N}}{2} = \frac{1}{2}\sum_{m=M+1}^{M+N}(n_{-m}-n_{m})\;,
\end{equation}
which is due to the same left-to-right current that we have examined in previous Sections, albeit no longer in a steady state. For clarity we return to the zero-temperature scenarios; the frequency smoothing effect of finite temperature can be added straightforwardly to the finite-size and finite-time effects that we now examine.

As we can see in Fig.~5), there is a brief initial period in which the fermions first begin to move through the middle chain of sites, via the weak links. Within a time that is at most of order $M/\tau$, however, this initial onset phase ends and we can see a remarkably steady fall of $\Delta\mathcal{N}$. The initial equilibrium state in the left end of the lattice can be viewed as an ensemble of wave packets moving in both directions; these packets move to and through the middle chain from further back in the left reservoir, just as they would if the reservoir were infinite. Even the packets which begin moving to the left, and then approach the middle chain after reflecting from the left end of the lattice, simply reproduce the effect of incoming packets from beyond the lattice end. The first effect which does not look just like an infinite system, in fact, is the return to the weak link of packets which have been reflected from it, and then reflected again from the left end of the lattice. This occurs for packets with group velocity $v_G$ at time $t=2N/v_G$. The maximum group velocity in our system is $2\tau$, occurring at $\omega =0$, and so the earliest finite size effects are those that appear in Fig.~5 at $t=N/\tau$. Between $t\gtrsim M/\tau$ and $t< N/\tau$ there can be plenty of time to observe a quasi-stationary current, as long as $N$ is sufficiently larger than $M$. And as Fig.~5 shows, the rate at which $\Delta\mathcal{N}$ falls, during this quasi-stationary epoch between the initial transient and the onset of finite-size effects, can closely match the steady-state current of the infinite system.

\begin{figure}\label{fig:quasi-steady0}
\includegraphics[width=.45\textwidth]{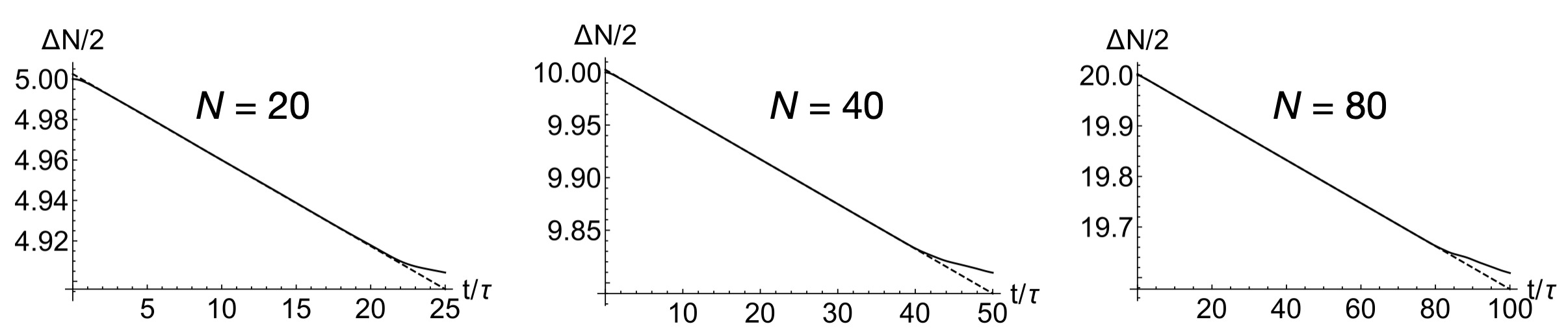}
	\caption{Finite-time evolution in finite systems. The initial state is prepared in a finite lattice with $N=20$, 40, or 80 sites in the outer lattice on each side. With the weak links between reservoir ends and middle chain initially turned off completely ($\tau_1\to 0$), the $N$-site left end lattice is initially half-filled at zero temperature, with the right end and central chain of seven sites ($M=3$) both initially empty. At $t=0$ the weak links are turned on $\tau_1\to 0.1$ and the system's single-particle density matrix is evolved in time under (\ref{closed}) numerically, for a duration $2.5 N/\tau$. The difference in total left- and right-side populations $\Delta N /2 = (N_L-N_R)/2$ is plotted in each case of $N$. The dashed line has the slope $-I$, where $I$ is the steady-state current for the $N\to\infty$ limit as shown in Fig.~3. The early transient epoch is so brief that it can only barely be seen with the shorter time axis of the $N=20$ plot. The onset of finite-size effects at $t=N/\tau$ is clear in all cases: with group velocity $2\tau$ for half filling, this is how long it takes for particles that initially reflect from the weak link to return to the weak link after reflecting again from the end of the lattice. For times after the short initial transient and before $t=N/\tau$, the finite system's particle difference closely reproduces the infinite system's steady-state current---in the case $\mu_L=0$ of initial half-filling. See Fig.~6 for other cases of $\mu_L$.}
\end{figure}

\subsection{Extended transient effects near transmission resonances}
In fact Fig.~5 provides an overly optimistic view of how easy it is to see steady-state current with finite-time measurements: although $\mu_L=0$ is the worst case for finite-size effects, since $v_G$ is highest there, it turns out to be the best case for short duration of the initial transient period. While $\mu_L=0$ is not so much better than typical $\mu_L$ that are not close to resonances, the slope of $\Delta\mathcal{N}$ can take longer to settle down steadily for $\mu_L$ near transmission resonances other than the one at $\mu_L=0$; see Fig.~6. The curves of $\Delta\mathcal{N}(t)$ appear quite similar for different outer-lattice sizes $N$, when both $t$ and $\mu_L$ are rescaled with $N$ appropriately, indicating that the extended transient epochs near transmission resonances are probably not really finite-size effects, but rather finite-time effects. We cannot avoid these problems by letting the system evolve for longer times, however, because finite-size effects will appear if we wait too long.

\begin{figure}\label{fig:twelve}
\includegraphics[width=.45\textwidth]{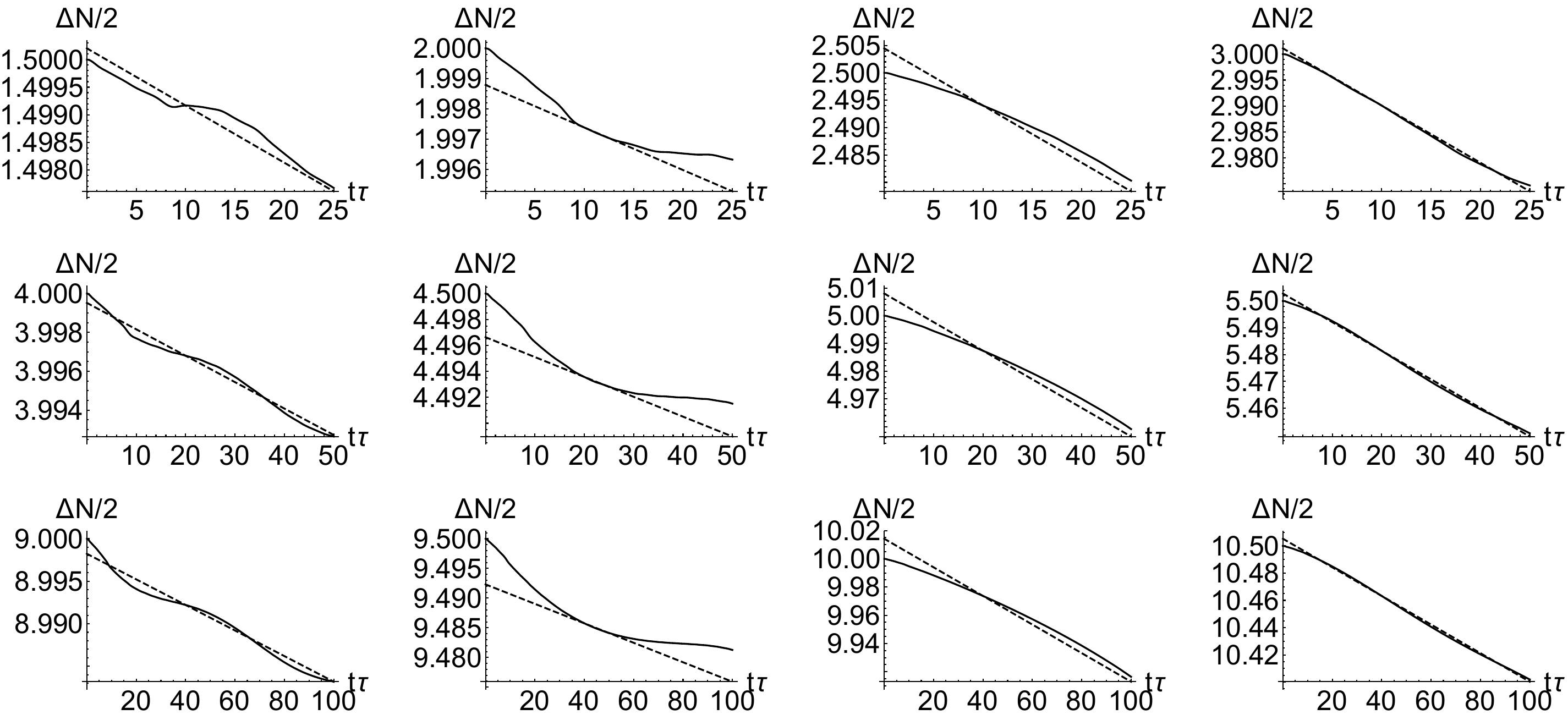}
	\caption{Zero-temperature fermion flow over time in finite systems with different $N$ as in Fig.~5, for $\mu_L$ near the first conductance step at $\mu_L=-\sqrt{2}$, and times less than $t=2.5 N$. Top row: $N=20$ sites in the lattice ends, with initial Fermion numbers 3,4,5,6 from left plot to right. Middle row: $N=40$, initial numbers 8,9,10,11. Bottom row: $N=80$, initial numbers $19,20,21,22$. The different initial particle numbers all correspond to $\mu_L$ near $-\sqrt{2}$; in particular the second column of plots have $\mu_L$ just below $-\sqrt{2}$ and the third column have $\mu_L$ just above it. Since the group velocities at frequencies near these $\mu$ are near $\sqrt{2}$ rather than 2 as at $\omega=0$, finite-size effects are not yet apparent. Finite-time effects, however, are clear: the initial transient epoch lasts much longer near the conductance resonance.}
\end{figure}

\subsection{Quasi-stationary current versus $\mu_L$}
While finite lattices may thus force finite-time problems upon us, though, these problems may not actually be too severe. Fig.~7 shows finite-system analogs of Fig.~3a), with three different cases of $N$ (20, 40, and 80 in Figs.~7a), 7b), and 7c), respectively). Each of these three plots further shows three different simple attempts to extract the quasi-steady current from the growth of $\Delta\mathcal{N}(t)$ before the onset of finite-size effects at $t=N/\tau$. These attempts represent $I$ as local slopes $[\Delta\mathcal{N}(t_m+1/\tau)-\Delta\mathcal{N}(t_m-1/\tau)]/(2\tau)$, for three different choices of the measurement time $t_m$: the early choice $t_m=N/(4\tau)$, the late choice $t_m=3N/(4\tau)$, and the intermediate choice $t_m=N/(2\tau)$. If $\Delta\mathcal{N}(t)$ were as straight as it is in Fig.~5, at $\mu_L=0$, these three choices of $t_m$ would all yield very similar $I$ estimates. Discrepancies between the three attempts to find $I$ show the non-stationary nature of the fermion flow at other $\mu_L$.
\begin{figure}\label{fig:quasi-steady1}
\includegraphics[width=.45\textwidth]{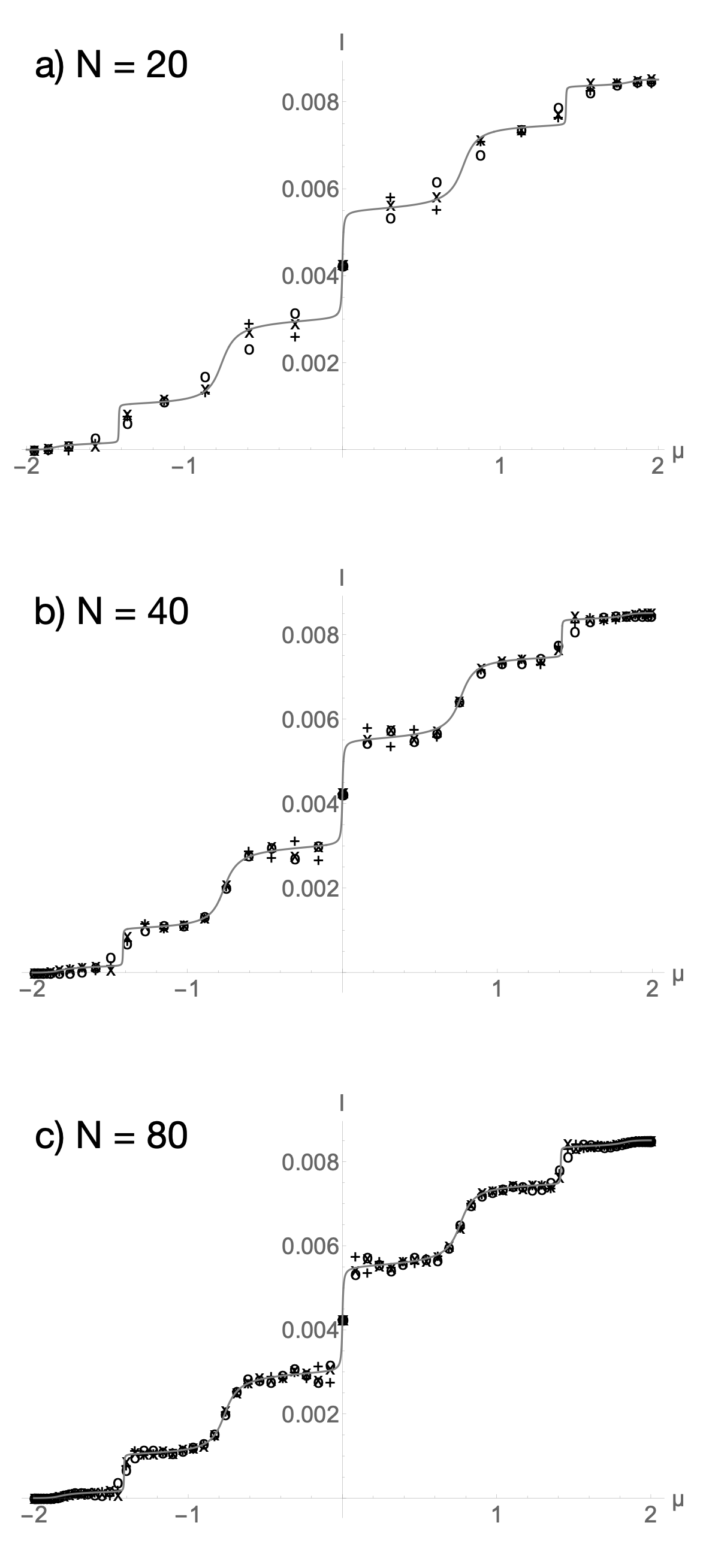}
	\caption{Zero-temperature current $I$ versus initial left-side chemical potential $\mu_L=\mu$ computed at three different times before the onset of finite-size effects, for each of three different finite lattice sizes. All cases have $\tau_1=0.1$, $\gamma=0.5$, $M=3$, and initial $\mu_R<-2$ (no particles in the right reservoir section). All three plots a), b), c) show the same thick, gray curve, which is $I(\mu)$ for the steady state of the infinite lattice with $\Lambda=1$ and $V=0$. In plots a), b), and c) we show cases with $N=$ 20, 40, 80. Each plot shows $I$ computed as $\big([n_R(t+1)-n_L(t+1)]-[n_R(t-1)-n_L(t-1)]/2$ for $t=0.25 N$ (o markers), $t=0.5 N$ (+ markers), and $t=0.75 N$ (x markers). The markers are discrete points rather than continuous lines, because fermions at zero temperature in a finite lattice simply occupy the $n$ lowest modes, for integer $n$ between 1 and $N$, and so the initial state only actually changes with chemical potential $\mu$ in discrete steps.}
\end{figure}

As Fig.~7 shows, it is possible to resolve the steady-state current $I(\mu_L)$ from Fig.~3a) fairly well with finite lattices as small as $N=20$, and lattices as large as $N=80$ allow essentially duplicating the steady-state current curve of the infinite system, with clear conductance steps that are noticeably smoother for the even-parity channels that are affected by loss. Even though extended transient effects near the steps still exist even for $N=80$, as shown in Fig.~6, this issue only really affects the two data points on either side of the step, and for $N=80$ there are many other points to map the curve out accurately. 

Ultimately the main limitation from finite lattices turns out to be the fundamentally limited frequency resolution from only having an $N$-dimensional single-particle Hilbert space in an $N$-site lattice. As long as $N$ is larger than $M$ by enough both to delay finite-size effects until the initial transient epoch is mostly over and to resolve the $2M+1$ conductance steps in the frequency band $-2<\omega<2$, experiments with ultracold Fermi gases in finite optical lattices can indeed reveal the same interesting transport phenomena with a lossy site that have been predicted for finite chains of quantum dots coupled to leads.

\section{Discussion and Outlook}

In conclusion, we have constructed a class of fermionic lattice gas models which can reproduce the results of \cite{AMTC} in the limit $\Lambda\to\infty$, and then represent quantum gas experiments in the limit $\Lambda\to 1$. Different but similar effects are predicted for quantum gases in optical lattice, in comparison with the parity-dependent conductance steps that are predicted to be seen in quantum dot chains between leads. Finite-temperature effects seem likely to be a practical challenge in near-future experiments with quantum gases. While these effects will not prevent observation of the most interesting conductance features, they will likely prevent quantum gas experiments from matching the sharpness of conductance steps that can be seen in condensed matter systems.

As in the condensed matter context of \cite{AMTC}, we have also focused in this paper on the symmetry dependence of loss effects, whereby localised loss smoothes out some conductance steps while leaving other steps sharp because their associated transport channel is protected from loss by symmetry. In particular we have considered loss at the central site only, so that channels with odd parity are unaffected. The flexibility of cold atom experiments may also open new avenues, however, to using tailored loss for more general control over quantum transport, such as by tuning loss at multiple sites in order to filter out more transport channels, or by making the loss rates time-dependent in order to select particular wave packets for transport while absorbing all others.

One entirely new possibility that can appear with quantum gases in lattices is transport of bosons instead of fermions. In the absence of interactions, however, bosons do not readily display the dramatic step-like conductance of zero-temperature fermions, because Bose-Einstein distributions cannot replicate the moveable sharp edge of a low-temperature Fermi-Dirac distribution with variable Fermi energy. For non-interacting bosons, the chemical potential can rise no higher than the bottom of the single-particle energy band; the only way to have varying exposure to the different discrete transport channels through the finite middle chain is to vary the reservoir temperatures. 
\begin{figure}\label{fig:bosons}
\includegraphics[width=.45\textwidth]{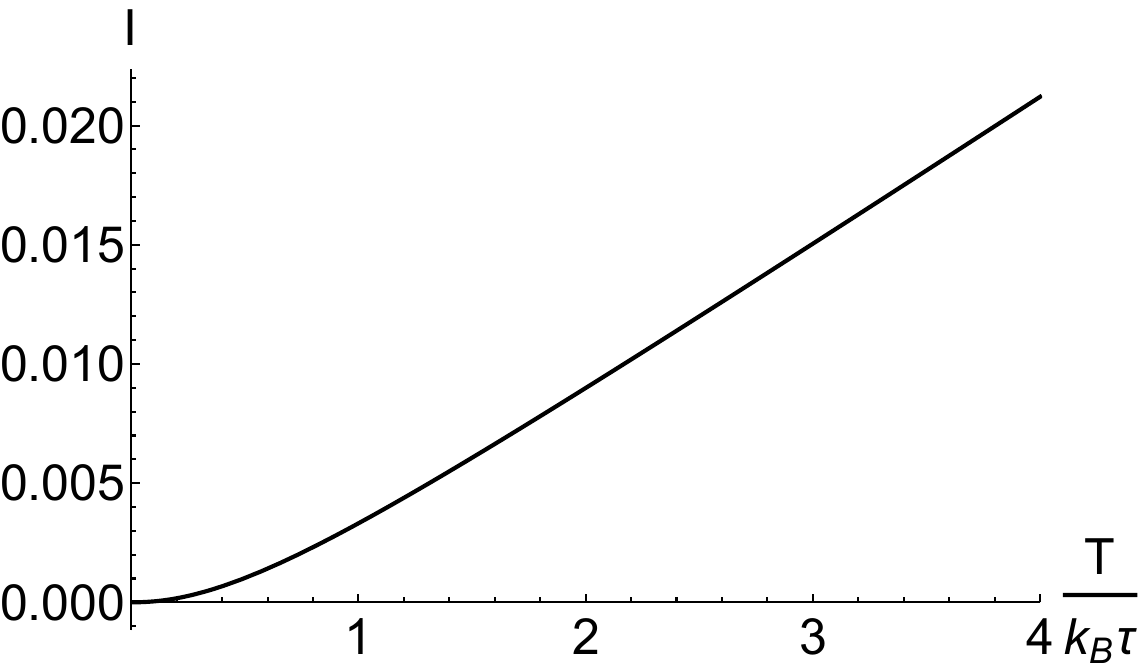}
	\caption{Current $I$ through the finite chain of seven sites ($M=3$) as in Fig.~3, but for non-interacting bosons at maximal chemical potential $\mu_L=-2$, as a function of temperature. No conductance steps are apparent: while the sharp edge of the zero-temperature Fermi-Dirac distribution rises with increasing $\mu$, increasing temperature in the Bose-Einstein distribution merely spreads the distribution out flatter, steadily overlapping more and more with all seven conductance resonances.}
\end{figure}

As Fig.~8 disappointingly shows, increasing temperature simply mixes in higher-frequency channels continuously. No conductance steps can be seen---at least not in scenarios like those we have considered in this paper, and not without interactions among the bosons. Perhaps different kinds of experiments will be able to show more abrupt unlocking of discrete transport channels for bosons. Further studies may also reveal more interesting transport behavior in interacting bosons, which can, among other complications, have higher chemical potentials.

\acknowledgements{
The author thanks Corinna Kollath and A.-M. Visuri for very useful discussions, and acknowledges support from the Deutsche Forschungsgemeinschaft (DFG) through SFB/TR185 (OSCAR), Project No.~277625399. In particular the essential ideas of representing lead contacts as weak links, and addressing finite-size effects, are due to C.~Kollath.}

\section{Appendix A: The single-particle wave functions $\Psi^{L,R}_m(\omega)$}
\subsection{Unnormalized eigenfunctions of definite parity for $|m|\leq M$}
For $V\not=0$ our entire system lacks reflection symmetry, but there is always reflection symmetry within the middle chain $|m|\leq M$. This makes it straightforward to construct $\psi_m$ eigenvectors that are either odd or even in $m$, for $|m|\leq M$. Once we have these solutions, we will be able to combine them into left-incident and right-incident $\Psi^{L,R}_m$, and finally normalize them. 

The odd solutions are the simplest, since they vanish at $m=0$ and are therefore unaffected by the imaginary potential term $\gamma$. They can simply be written as
\begin{align}\label{psim}
    \psi^-_m(\omega) &\underset{|m|\leq M}{=} \sin\big(k(\omega)m\big)\nonumber\\
    \omega &=-2\cos(k)\nonumber\\
    k(\omega) &= \cos^{-1}\Big(-\frac{\omega}{2}\Big)\;.
\end{align}
Note that we may need to consider $|\omega|>2$, for which $k(\omega)$ is complex, because for $\Lambda>1$ the outer parts of the lattice support eigenmodes in a wider bandwidth than that of the middle chain. 

The modes that are even in the middle chain are a bit less trivial, because of the complex potential at $n=0$:
\begin{align}
    \psi^+_m(\omega) &\underset{|m|\leq M}{=} \cos\big(k(\omega)m\big) - \frac{i\gamma}{4\sin\big(k(\omega)\big)}\sin\big(k(\omega)|m|\big)\;.
\end{align}

How these eigenvectors behave for $|m|>M$ is then determined firstly by solving (\ref{1PS}) for $|m|>(M+1)$, which implies
\begin{align}\label{DOS2}
    \psi_n^\pm(\omega) &\underset{|m|>M}{=} \alpha^\pm_{L,R} e^{ik_{L,R}(\omega)|m|} + \beta^\pm_{L,R} e^{-ik_{L,R}(\omega)|m|}\;,
\end{align}
for $|m|>M$, with the subscript $L$ applying for $m<-M$ and $R$ for $m>M$, and with $k_{L,R}(\omega)$ defined in (\ref{DOS1}) above. 

To fix the separate left and right coefficients $\alpha^\pm_{L,R}$ and $\beta^\pm_{L,R}$, we finally solve (\ref{1PS}) for $|m|=M$ and $|m|=M+1$, which provide two conditions at each end of the finite middle chain that are analogous to the continuity conditions on the wave function and its first derivative in continuous Schr\"odinger problems. After some algebra we obtain
\begin{align}\label{parity}
    \psi_m^- \underset{|m|>M}{=}& \mathrm{sgn}(m)\Big[\frac{A_{L,R}}{2}e^{ik_{L,R}|m|}+\frac{A^*_{L,R}}{2}e^{-ik_{L,R}|m|} \Big]\nonumber\\
    \psi_m^+ \underset{|m|>M}{=}& \Big(B_{L,R}-\frac{i\gamma}{4\sin(k)}A_{L,R}\Big)\frac{e^{ik_{L,R}|m|}}{2}\nonumber\\
    &+ \Big(B^*_{L,R}-\frac{i\gamma}{2
    4\sin(k)}A^*_{L,R}\Big)\frac{e^{-ik_{L,R}|m|}}{2}
\end{align}
where the $L$ subscripts apply for $m<-M$ and the $R$ subscripts for $m>M$. The coefficients here are
\begin{align}\label{AB}
A_{L,R}(\omega) =& \frac{e^{-ik_{L,R}M}}{i\sin(k_{L,R})}\nonumber\\
&\times\left(\frac{1}{\tau_1}\sin[k(M+1)]-\tau_1\sin(kM)e^{-ik_{L,R}}\right)\nonumber\\
B_{L,R}(\omega) =& \frac{e^{-ik_{L,R}M}}{i\sin(k_{L,R})}\nonumber\\
&\times\left(\frac{1}{\tau_1}\cos[k(M+1)]-\tau_1\cos(kM)e^{-ik_{L,R}}\right),
\end{align}
where $k$ without $L,R$ subscripts is the middle-chain $k(\omega)$ from (\ref{kom}), which is reproduced in (\ref{psim}).
Although we have the same $\Lambda$ for both $m>M$ and $m<-M$, if $V\not=0$ then $k_L\not=k_R$ and so $\psi_m^\pm$ do not actually have definite parity for $|m|>M$.

\subsection{Left- and right-incident wave functions}
We then simply construct linear combinations of $\psi^\pm_m$ which contain no incoming wave on one side, and in which the amplitude of the incoming wave on the other side is $1/\sqrt{2\pi}$. The result is
\begin{align}\label{LRZ}
    \Psi^L_m(\omega) &= \frac{A^*_R\psi_m^+(\omega)-\Big(B_R^*-\frac{i\gamma}{4\sin(k)}A^*_R\Big)\psi_m^-(\omega)}{\sqrt{2\pi}Z}\nonumber\\
    \Psi^R_m(\omega) &= \frac{A^*_L\psi_m^+(\omega)+\Big(B^*_L-\frac{i\gamma}{4\sin(k)}A^*_L\Big)\psi_m^-(\omega)}{\sqrt{2\pi}Z}\nonumber\\
    Z(\omega)&=\frac{A^*_RB^*_L+A^*_LB_R^*}{2}-\frac{i\gamma}{4\sin(k)}A^*_LA^*_R
\end{align}
which further implies
\begin{align}\label{TR}
    \mathcal{T} &= \frac{i}{Z}\frac{\sin(k)}{\sqrt{\sin(k_L)\sin(k_R)}}\nonumber\\
    \mathcal{R}_L &= \frac{1}{2Z}\left(A^*_R B_L + B^*_R A_L -\frac{i\gamma}{2\sin(k)}A^*_RA_L\right)\nonumber\\
    \mathcal{R}_R &= \frac{1}{2Z}\left(A^*_L B_R + B^*_L A_R -\frac{i\gamma}{2\sin(k)}A^*_LA_R\right)\;.
\end{align}
As we note in Section III of our main text, the dimensionless parameter $\Lambda$ does not appear explicitly in the results for $\mathcal{R}_{L,R}$ and $\mathcal{T}$. $\Lambda$ does appear in Equation (\ref{DOS1}), however, which defines $k_{L,R}(\omega)$, and so $\Lambda$ actually does affect $\mathcal{R}_{L,R}$ and $\mathcal{T}$ substantially. In particular both for $\Lambda=1$ and for $\Lambda\to\infty$ we obtain $k_L = k_R$ and the distinctions between $A_{L,R}$, $B_{L,R}$ disappears, so that we can take $\mathcal{R}_{L,R}\to\mathcal{R}$ as in our main text.

\subsection{Infinite $\Lambda$ limit}
As we note in Section IV of our main text, the $\Lambda\to\infty$ limit which makes the reservoir density of states constant over the middle chain bandwidth lets us replace $k_{L,R}\to\pi/2$. Our expressions for $Z$, $\mathcal{T}$, and $\mathcal{R}_{L,R}$ then simplify greatly. We obtain\begin{widetext}
\begin{align}
A_{L,R}\underset{\Lambda\to\infty}{\longrightarrow}& (-i)^{M+1}\left(\frac{\sin[k(M+1)]}{\tau_1}+i\tau_1\sin(kM)\right)\nonumber\\
B_{L,R}\underset{\Lambda\to\infty}{\longrightarrow}& (-i)^{M+1}\left(\frac{\cos[k(M+1)]}{\tau_1}+i\tau_1\cos(kM)\right)\nonumber\\
|Z|^2 \underset{\Lambda\to\infty}{\longrightarrow}& \left[\frac{\sin[k(M+1)]\cos[k(M+1)]}{\tau_1^2}+\tau_1^2\sin(kM)\cos(kM)\right]^2\nonumber\\
&+ \left[\sin(k)+\frac{\gamma}{4\sin(k)}\left(\frac{\sin^2[k(M+1)]}{\tau_1^2} + \tau_1^2\sin^2(kM)\right)\right]^2\nonumber\\
|\mathcal{T}|^2\underset{\Lambda\to\infty}{\longrightarrow}& \frac{\sin^2(k)}{|Z|^2}\nonumber\\
|\mathcal{R}_{L,R}|^2\underset{\Lambda\to\infty}{\longrightarrow}&\frac{1}{|Z|^2}\left[\left(\frac{\sin[k(M+1)]\cos[k(M+1)]}{\tau_1^2}+\tau_1^2\sin(kM)\cos(kM)\right)^2\right.\nonumber\\
&+\left.\frac{\gamma^2}{16\sin^2(k)}\left(\frac{\sin^2[k(M+1)]}{\tau_1^2}+\tau_1^2\sin^2(kM)\right)^2\right]\;,
\end{align}\end{widetext}
as used in our main text.

\section{Appendix B}
As we noted in Section III of our main text, for the class of models that we have defined it can in general happen that one of $k_{L,R}$ is complex---or even that both of them are.
This can happen for one of $k_{L,R}$, for $V>0$, because the bandwidth $4\Lambda\tau$ of the outer lattice ends is finite, and so a relative shift of $V$ between the bands of the outer ends means that they do not fully overlap. Wherever an $\omega$ is in the continuous part of the infinite model's spectrum, because it is in the band of one lattice end, $k_{L,R}$ will be purely imaginary for $\omega<-2\Lambda \pm V/2$, below the corresponding lattice end's frequency band, and it will be $\pi$ plus an imaginary term for $\omega>2\Lambda \pm V/2$, above the band.

It can also happen that both $k_{L,R}$ are complex, for a discrete set of complex $\omega$ that are in the spectrum of our non-Hermitian Hamiltonian which includes the imaginary potential at $m=0$, if $\Lambda<1$ so that the middle chain possesses eigenstates outside the bands of both the outer lattice ends. 

We have only considered $\Lambda\geq 1$ in this paper, leaving other cases for future work. In fact we have only explicitly considered cases either with $V>0$ but $\Lambda\to\infty$, so that $k_{L,R}$ are both real for all $\omega$ that are involved in transport through the middle chain, or else with $\Lambda=1$ but $V=0$, so that all $k_{L,R}$ are real. For the benefit of future work, however, we note that our formulas extend straightforwardly to include complex $k_{L,R}$, when they are taken in the general forms shown in Appendix A above, and when the observables include the $X_{L,R}$ factors as in Appendix C below.

If for example $k_R(\omega)$ has an imaginary part for some $\omega$ but $k_L(\omega)$ is real, we can keep our $\Psi_n^L(\omega)$ as described above, with the stipulation that we take the positive imaginary branch of $k_R$, but there will simply be no $\Psi_n^R$ for this $\omega$, because the ``incoming" wave would be exponentially growing. It will also be important to note that the evanescent transmitted waves in these cases of $\Psi_n^L(\omega)$ carry no particle or energy currents to $n\to+\infty$ even though $|\mathcal{T}|>0$.

\section{Appendix C}
\subsubsection{Particle current with complex $k_{L,R}$}
In the general case where $k_{L,R}$ are not equal, and may be complex, it is necessary to modify our main text's formula (\ref{Iint}) for the particle current through the middle chain, by distinguishing $\mathcal{R}_L$ and $\mathcal{R}_R$, and by including additional factors $X_{L,R}(\omega)$:
\begin{align}\label{Iint2}
    I =& \frac{1}{4\pi\tau_1^2}\int\!d\omega\,\left[f_L(\omega)\Big(1+X_L|\mathcal{T}|^2-|\mathcal{R}_L|^2\Big)\right.\nonumber\\
    &
    \quad\quad\quad\quad-\left.f_R(\omega)\Big(1+X_R|\mathcal{T}|^2-|\mathcal{R}_R|^2\Big)\right]\;.
\end{align}
The factor $X_{L}$ here is equal to one if $k_{R}$ is real, and is zero if $k_R$ has a non-zero imaginary part; $X_R$ depends in the same way on whether $k_L$ is purely real. These factors do not have to be inserted by hand, but come out simply in evaluating $J_m$ in the cases where $k_L$ or $k_R$ is either imaginary or imaginary plus $\pi$.

The subscripts are not accidentally switched: the factor $X_L$ depends on $k_R$, and \emph{vice versa}, because the issue is whether or not the wave which is transmitted through to the right from the left, or \emph{vice versa}, is evanescent or propagating. 

\subsubsection{Local energy at site $m$}
The Hamiltonian (\ref{H}) is a sum over sites $m$, and apart from site $m=0$, where there is an imaginary potential, each term in the $m$-sum in (\ref{H}) is an observable that can be interpreted as the local energy associated with site $m$; we can treat each site's two neighbors symmetrically by averaging their contributions. The expectation value of this local energy is then
\begin{align}
    E_m =& \hbar\tau \Big(-\frac{T_m}{2} \big(R_{m+1,m}+R_{m,m+1}\big)\nonumber\\
    &-\frac{T_{m-1}}{2} \big(R_{m,m-1}+R_{m-1,m}\big)+V_m R_{mm}\Big)
\end{align}
according to (\ref{H}).

\subsubsection{Energy current}
Just as the particle number current $J_m$ could be inferred from the time derivative of the average local number $\bar{n}_m$, we can also compute from (\ref{closed}) that
\begin{align}
    \frac{d}{dt}E_m &= - (Q_{m+1}-Q_m) -\frac{\gamma\tau}{2}E_0 (\delta_{m,-1}+\delta_{m0})\nonumber\\
    Q_m &= -i\frac{\hbar\tau^2}{2}T_{m-1}\Big[T_m(R_{m+1,m-1}-R_{m-1,m+1})\nonumber\\
    &\qquad+T_{m-2}(R_{m,m-2}-R_{m-2,m})\nonumber\\
    &\qquad-(V_m+V_{m-1})(R_{m,m-1}-R_{m-1,m})\Big]\;,
\end{align}
and recognize $Q_m$ as the local energy current.
In the steady state $\dot{E}_m = 0$ then implies that $Q_m$ is piecewise constant,
\begin{equation}
    Q_m = Q -\frac{\gamma\tau}{2}\,\mathrm{sgn}(m)E_0\;,
\end{equation}
with $Q_0 = Q$ also being the average energy current through the finite middle chain from left to right. 

Evaluating $Q = \lim_{m\to\infty}(Q_{-m}+Q_m)/2$ in (\ref{Rmmstat}) and using (\ref{1PS}) then reveals a result which we might well have guessed without even thinking much about what it meant: the average energy current in the stationary state is a similar integral to the integral (\ref{Iint}) for the average particle current $I$, with the integrand simply multiplied by $\hbar\tau\omega$.
\begin{align}\label{Qint}
    Q =& \frac{\hbar\tau^2}{4\pi}\int\!d\omega\,\omega\left[f_L(\hbar\tau\omega)\Big(1+X_L|\mathcal{T}|^2-|\mathcal{R}_L|^2\Big)\right.\nonumber\\
    &\qquad\left.-f_R(\hbar\tau\omega)\Big(1+X_R|\mathcal{T}|^2-|\mathcal{R}_R|^2\Big)\right]\;.
\end{align}

\section{Appendix D}
\subsection{Orthonormality and completeness of solutions to Equation (\ref{1PS})}
Our single-particle Schr\"{o}dinger equation (\ref{1PS}) is non-Hermitian, because of its imaginary potential term, and so the usual orthogonality among its eigenfunction solutions does not hold. We therefore cannot use the usual form of orthonormality to show completeness of the eigenfunctions as a basis for all initial states.

We do, however, have orthogonality in a generalised form. For any solution $\psi_m(\omega)$ to (\ref{1PS}), we define the dual function
\begin{equation}
    \bar{\psi}_m(\omega) = \psi_m^*(\omega^*)\Big|_{\gamma\to-\gamma}\;.
\end{equation}
Since $\bar{\psi}_m$ thus satisfies (\ref{1PS}) just as $\psi_m$ itself does, we can deduce from (\ref{1PS}) that
\begin{align}
    (\omega-\omega')\bar{\psi}_m(\omega')\psi_m(\omega) &= -T_m\big(\bar{\psi}_m\psi_{m+1}-\bar{\psi}_{m+1}\psi_m\big)\nonumber\\
    &\quad- T_{m-1}\big(\bar{\psi}_m\psi_{m-1}-\bar{\psi}_{m-1}\psi_m\big)\nonumber\\
    &\equiv X_m - X_{m-1}
\end{align}
so that 
\begin{align}
    (\omega-\omega')\sum_{m=-L}^L\bar{\psi}_m(\omega')\psi_m(\omega) =
X_L-X_{-L-1}
\end{align}
which for $L\to\infty$ is an infinitely rapidly oscillating function of $\omega$ and $\omega'$ and thus zero as a distribution.

Normalising these dual functions $\bar{\psi}_m(\omega)$ appropriately, we can then construct a partition of unity,
\begin{equation}\label{Po1}
    \int\!d\omega\,\bar{\psi}_{m'}(\omega)\psi_m(\omega) = \delta_{m'm}\;.
\end{equation}
It is clear that we can do this for $\gamma=0$, since then the system is Hermitian, and it is also clear from the form of our $\Psi^{L,R}_m$ that for any $\gamma>0$ we can construct arbitrary wave packets at infinity out of our $\psi_m$. The only way that (\ref{Po1}) can at some point fail, as we continuously raise $\gamma$ from zero, is for a new $\psi_m$ to appear in the spectrum, which must be a bound state in the sense that it vanishes as $|m|\to\infty$. If this happens, as for example it does at $\gamma = 2$ for $\Lambda=\tau_1 =1$, then we must simply add the additional states to the $\omega$ spectrum, so that the integral in (\ref{Po1}) must be supplemented with a sum over the discrete bound $\psi_m$. 

Given (\ref{Po1}) and orthonormality with the $\bar{\psi}_m$ dual states, we simply need to add any necessary discrete states to the integrals in (\ref{Rmmt}) (and let $\omega'\to\omega'^{*}$ for the complex discrete $\omega$), while choosing
\begin{equation}A_{j'j}(\omega,\omega') = e^{i\tau(\omega-\omega'^*)t_I}\sum_{l,l'}R_{l',l}(t_I)\bar{\psi}_l(\omega)\bar{\psi}^*_{l'}(\omega'^*)\end{equation}
to ensure that $R_{m'm}(t)$ as given by (\ref{Rmmt}) satisfies arbitrary initial conditions at $t=t_I$.

In fact no discrete bound state $\psi_m$ can be involved in any of the stationary $R_{m'm}$ that we actually consider in this paper, because the bound states must indeed all have complex $\omega$ (allowed for the non-Hermitian Schr\"{o}dinger equation (\ref{1PS})), with imaginary parts such that they decay exponentially in time. 

\end{document}